\pdfoutput=1

\documentclass[11pt]{article}

\usepackage[preprint]{acl}

\usepackage{times}
\usepackage{latexsym}

\usepackage[T1]{fontenc}

\usepackage[utf8]{inputenc}

\usepackage{microtype}

\usepackage{inconsolata}

\usepackage{graphicx}

\usepackage[utf8]{inputenc} 
\usepackage[T1]{fontenc}    
\usepackage{hyperref}       
\usepackage{url}            
\usepackage{booktabs}       
\usepackage{amsfonts}       
\usepackage{nicefrac}       
\usepackage{microtype}      
\usepackage{xcolor}         

\usepackage{multirow}
\usepackage{graphicx}
\usepackage{ulem}
\usepackage{caption}
\usepackage{subcaption}
\usepackage{amssymb}
\usepackage{booktabs}
\usepackage{multibib}
\usepackage{colortbl}
\usepackage{makecell}
\usepackage{wrapfig}
\usepackage{color}
\usepackage{float}
\usepackage{inconsolata}
\usepackage{amsmath}

\usepackage{enumitem}
\setlist{leftmargin=1em}

\usepackage[listings]{tcolorbox}
\tcbuselibrary{listings,theorems}
\newtcolorbox{mybox}{colback=white!5!white,colframe=black!75!black, left=.05in, right=.05in}

\definecolor{bluex}{rgb}{0.27, 0.42, 0.81}
\definecolor{purplex}{HTML}{9564bf}
\definecolor{red3}{HTML}{C52A20}
\definecolor{red2}{HTML}{B36A6F}
\definecolor{red1}{HTML}{FFb5b5}
\definecolor{purple}{HTML}{B36A6F}
\definecolor{darkyellow}{HTML}{D5BA82}
\definecolor{blue1}{HTML}{508AB2}
\definecolor{blue2}{HTML}{C4E4E3}
\definecolor{green1}{HTML}{A1D0C7}
\definecolor{green2}{HTML}{BFF6BA}
\definecolor{green3}{HTML}{028100}
\definecolor{teal}{HTML}{508AB2}
\definecolor{purple1}{HTML}{8d3a94}

\newtcbtheorem[]{exmp}{Template}%
{colback=purple1!5,colframe=purple1!80,fonttitle=\bfseries, left=.02in, right=.02in,bottom=.02in, top=.02in}{exmp}

\title{Repository Structure-Aware Training Makes SLMs Better Issue Resolver}

\author{Zexiong Ma\thanks{\, Work done during the internship at Microsoft.}\hspace{0.4mm} $^{\diamondsuit,\clubsuit}$,\, Shengnan An$^{*\heartsuit,\clubsuit}$,\, Zeqi Lin\thanks{\, Corresponding authors.}\hspace{0.4mm} $^{\clubsuit}$,\\\textbf{Yanzhen Zou}$^{\diamondsuit}$,\,  \textbf{Bing Xie}$^{\dagger\diamondsuit}$
\vspace{1mm}\\
  $^{\diamondsuit}$School of Computer Science, Peking University,\,\,
  $^{\clubsuit}$Microsoft,\, 
  $^{\heartsuit}$Xi’an Jiaotong University\vspace{1mm}\\
  $^{\diamondsuit}$\texttt{mazexiong@stu.pku.edu.cn, \{zouyz, xiebing\}@pku.edu.cn}, \\
  $^{\heartsuit}$\texttt{an1006634493@stu.xjtu.edu.cn}, $^{\clubsuit}$\texttt{Zeqi.Lin@microsoft.com}
}

\begin{document}
\maketitle
\begin{abstract}
Language models have been applied to various software development tasks, but the performance varies according to  the scale of the models. Large Language Models (LLMs)  outperform Small Language Models (SLMs) in complex tasks like repository-level issue resolving, but raise concerns about privacy and cost. In contrast, SLMs are more accessible but under-perform in complex tasks. In this paper, we introduce \textbf{ReSAT} (Repository Structure-Aware Training), construct training data based on a large number of issues and corresponding pull requests from open-source communities to enhance the model's understanding of repository structure and issue resolving ability. We construct two types of training data: (1) \textbf{localization training data}, a multi-level progressive localization data to improve code understanding and localization capability; (2) \textbf{code edit training data}, which improves context-based code editing capability. The evaluation results on SWE-Bench-verified and RepoQA demonstrate that ReSAT effectively enhances SLMs' issue-resolving and repository-level long-context understanding capabilities.
\end{abstract}

\section{Introduction}

Language Models (LMs) have been applied to various software development tasks (such as code completion~\cite{Copilot}, code generation~\cite{chen2021evaluating,li2022competition, ma2024compositional, jiang2024survey}, and program repair~\cite{jin2023inferfix}), and many LM-based s have been integrated into real-world development processes. Language models can be categorized into two types based on their scales: (1) \textbf{Large Language Models (LLMs)}, with a parameter size of 100B or more, which are typically commercial, closed-source models~\cite{chatgpt, Claude, gpt4}.  (2) \textbf{Small Language Models (SLMs)}, with a parameter size of 13B or less, which are usually open-source models~\cite{codeqwen1.5, guo2024deepseek, team2024gemma, touvron2023llama, lozhkov2024starcoder}.  LLMs have stronger abilities but are usually only accessible via API, raising concerns about privacy leakage. SLMs, on the other hand, can be deployed on consumer-grade GPUs ~\cite{team2024gemma} but have poorer performance in more realistic, repository-level software development tasks (i.e., issue resolving~\cite{jimenez2024swebench}). 

Due to the poor performance of SLMs on repository-level tasks~\cite{liu2023repobench}, mainstream repository-level automatic programming assistants~\cite{devin} all employ  LLMs. 
SWE-Bench~\cite{jimenez2024swebench} is a benchmark to evaluate the ability of automatic programming assistants to resolve issues. Many works have employed agent-based or pipeline-based systems to utilize LMs for repository-level issue resolving, and all the top-performing works on the SWE-Bench leaderboard employ  LLMs \cite{gpt4, Claude}. 

Issue resolving requires language models to understand repository-level code and model long-range dependency. Given a lengthy code context and an issue, LMs are required to locate the code segment relevant to the issue, and generate code edits for the segments. Due to the limitations of model size and training data volume, neither the pre-training nor the instruction-tuning~\cite{luo2023wizardcoder} phases have endowed SLMs with the capability to perform such complex tasks effectively. There are rich structural information in open-source repositories, which is not utilized during training process of SLMs. 
Consequently, a pressing research question arises: \textit{Could we leverage the structural information in open-source repositories to enhance the repository-level understanding and issue-resolving capabilities of SLMs?}

\begin{figure*}
    \centering
    \subfloat[RAG-SWE first retrieves the most similar files to the issues, then generates code edits to the retrieved files.]{
    \label{fig:motivation1} 
    \includegraphics[width=.85\textwidth]{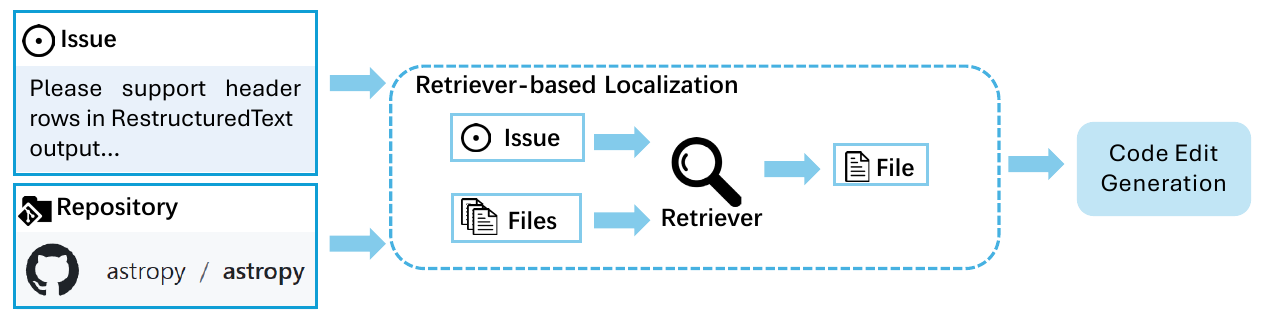}}\quad
    \subfloat[Agentless first performs LLM-based step-by-step progressive localization, then edits the localized code snippets.]{
    \label{fig:motivation2} 
    \includegraphics[width=.85\textwidth]{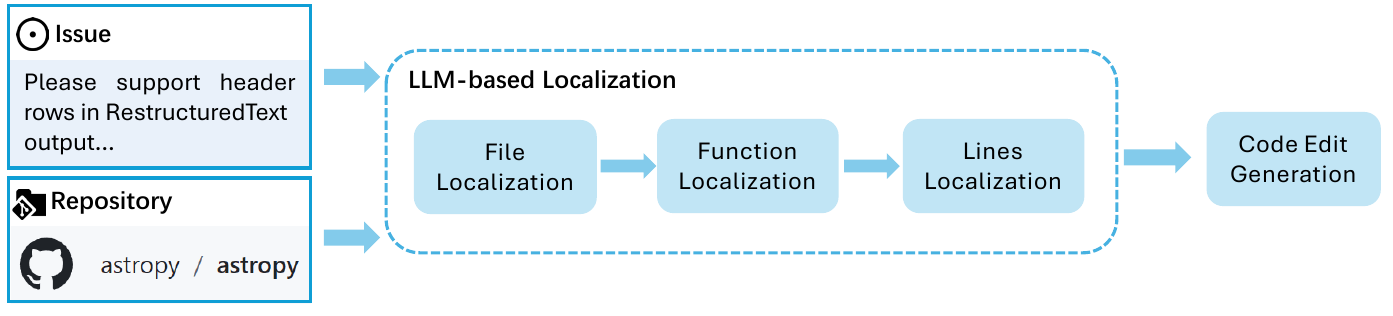}}\quad
    \caption{Typical issue resolving frameworks, both of them first localize the code snippets related to the issue, then edit the snippets to resolve the issue.}
    \label{fig:assistant}
\end{figure*}

In this paper, we propose \textbf{Re}pository \textbf{S}tructure-\textbf{A}ware \textbf{T}raining (\textbf{ReSAT}), to enhance the repository-level code understanding and issue resolving capabilities of SLMs. We crawl popular open-source projects from GitHub and the resolved issues within these projects, using them to construct localization and code edit training data. 
(1) We construct \textbf{localization training} data to improve the repository-level code understanding and feature localization abilities of SLMs. Based on different levels of localization granularity, we create three types of localization data: file-level, function-level, and line-level localization. 
(2) We also construct \textbf{code edit training} data to improve the SLMs' context-based code editing ability.

We utilize ReSAT to fine-tune two SLMs: \textit{CodeQwen1.5-7B-Chat}~\cite{codeqwen1.5} and \textit{Deepseek-Coder-6.7B-Instruct}~\cite{guo2024deepseek}, and apply them to two issue resolving frameworks: Agentless ~\cite{xia2024agentless} and RAG-SWE~\cite{jimenez2024swebench}. We utilize SWE-Bench-verified to evaluate the impact of ReSAT on the issue resolving capabilities of SLMs. Additionally, we assess the impact of ReSAT on the repository-level long-context understanding capabilities of SLMs through the RepoQA~\cite{liu2024repoqa} benchmark. We also conduct ablation studies to verify the effectiveness of the two parts of ReSAT.

In summary, this paper makes the following main contributions: 1) we propose ReSAT, a novel repository structure-aware training data construction approach, to improve the repository-level code understanding and issue resolving capibility of SLMs; 2) we apply ReSAT on two open-source SLMs, the experiments on SWE-Bench-verified and RepoQA benchmarks indicate the effectiveness of ReSAT; 3) we conduct analysis on the  unresolved issues of ReSAT-trained SLMs, providing guidance for further improving SLMs' issue resolving capabilities.


\section{Background}\label{sec:background}

In this section, we will provide a brief introduction to SWE-Bench, and introduce two existing issue resolving frameworks.


\subsection{SWE-Bench}

SWE-Bench~\cite{jimenez2024swebench} is a benchmark designed to test the capabilities of language models (LMs) in solving real-world software engineering problems. It consists of tasks derived from real GitHub issues and their corresponding pull requests across 12 popular python repositories. LMs or LM-based programming assistants are provided with an issue description and a repository, and expected to generate code edits to the repository that resolve the issue. Each issue is associated with test cases that can be executed in a Docker environment, and the evaluation is based on whether the edited repository can pass the test cases. SWE-Bench differs from traditional code generation benchmarks~\cite{chen2021evaluating, peng2024humaneval, austin2021program} by focusing on realistic software engineering tasks that require repository-level code understanding, making edits across multiple locations, and applying long-context reasoning.


SWE-Bench-verified~\cite{swebench-verified} is an updated benchmark proposed to address the evaluation problems in SWE-Bench. The tasks in SWE-Bench are derived from real GitHub issues, which introduces various problems, such as overly specific unit tests, vague problem descriptions, and complex environment setup requirements. To address these problems, OpenAI collaborated with the SWE-Bench team to create SWE-Bench-verified, an updated and more reliable issue resolving benchmark. 
All issue resolving experiments in this paper are conducted on SWE-Bench-verified.

\subsection{Issue Resolving Framework}

The intelligence level of GPT-4 makes it feasible to develop repository-level automatic programming assistants~\cite{wang2024opendevin, bairi2024codeplan}. Since the release of GPT-4, many efforts have been made to build issue resolving frameworks. 
Some approaches design framework that align with the software development workflow. Issue resolving framework divide complex task into subtasks, sequentially invoking LMs to complete different subtasks in order to achieve repository-level programming. \citet{jimenez2024swebench} propose the RAG-SWE framework. As shown in Figure~\ref{fig:motivation1}, RAG-SWE first utilizes a retriever to retrieve the files most similar to the issue description, then uses the retrieved files and the issue description as a prompt to make LMs generate code edits to resolve issues.
\citet{xia2024agentless} propose Agentless, which does not use a retriever, but instead allows LMs to directly locate the relevant parts based on the issue description and the structure of the repository.
In this setup, the LMs are responsible for identifying the location of the issue within the repository and generating candidate fixes. As shown in Figure~\ref{fig:motivation2}, Agentless first performs LLM-based step-by-step progressive localization, then edits the localized code snippets to resolve the issues. 
In this paper, we employ Agentless and RAG-SWE as our inference frameworks during the experiments.

\begin{figure*}
     \centering
     \small
     \includegraphics[width=0.9\linewidth]{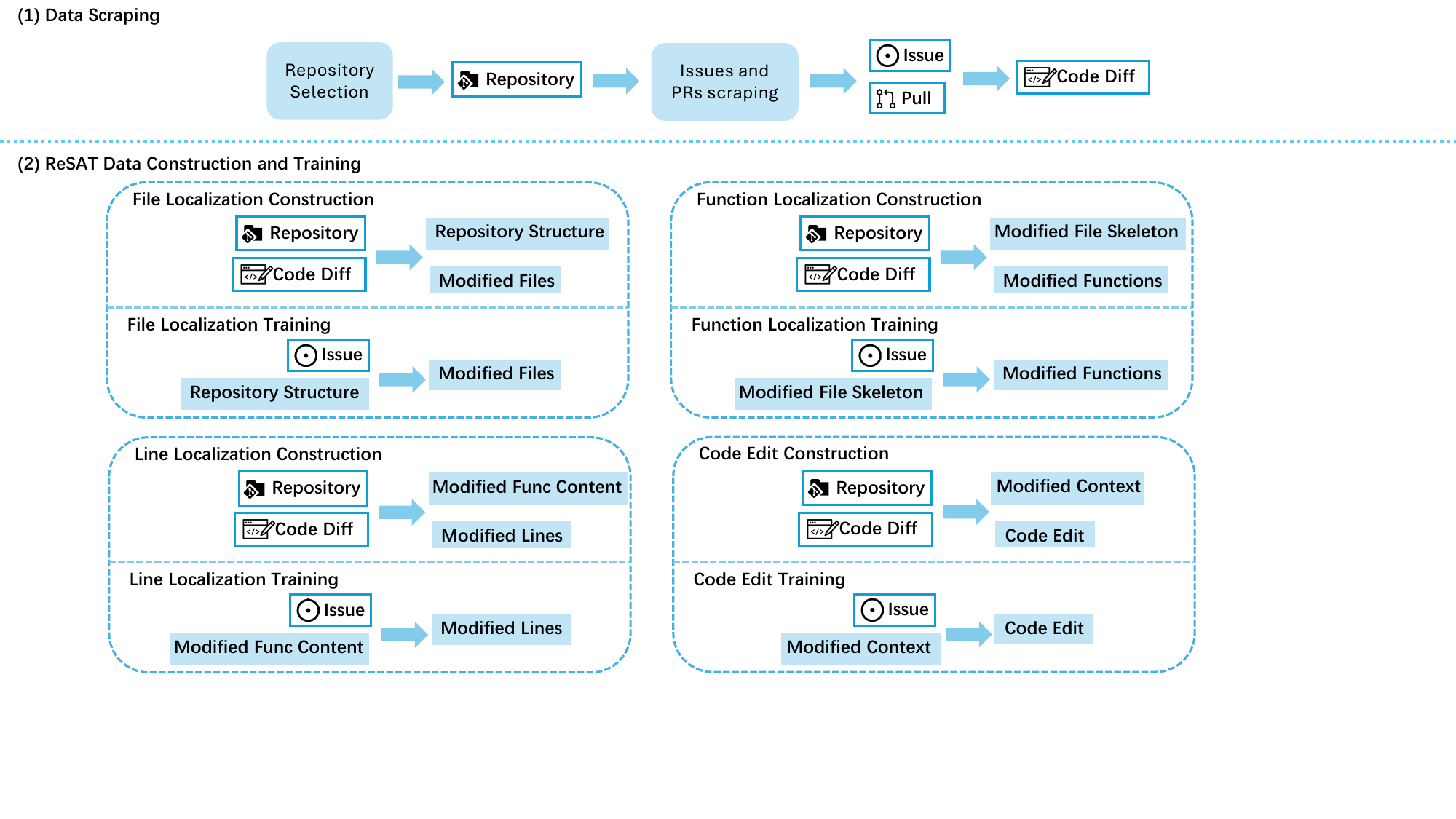}
     \caption{ReSAT training data scraping, construction and training pipeline. We first scrape issues and PRs from open-source repositories, then construct ReSAT localization and code edit data, and apply ReSAT training on SLMs.}
     \label{fig:approach}
 \end{figure*}

\section{Approach}\label{sec:approach}

In this section, we will introduce the details of Repository Structure-Aware Training, including data scraping, localization data construction, code edit data construction and training details.

\subsection{Data Scraping}
We leverage open-source repositories to construct structure-aware training data, addressing SLMs' lack of repository-level code understanding and issue resolving ability.
As shown in Figure ~\ref{fig:approach}(1), to construct ReSAT training data, we first select high-quality open-source repositories based on their download numbers, star count and Pull Request (PR) count. Then we scrape the resolved issues and corresponding PRs from the selected repositories.

\textbf{Repository Selection}. We select the most popular and active open-source python projects to scrape data and construct the ReSAT training dataset. Following previous work~\cite{jimenez2024swebench}, we assume that package quality is positively correlated with the number of downloads, scrape the top 5,000 most downloaded PyPI packages from the Top-pypi-packages website~\footnote{https://hugovk.github.io/top-pypi-packages/}, and filter out packages that do not contain corresponding GitHub repository on PyPI webset or without licenses that allow for free software use. To avoid data leakage, we exclude repositories that appeared in the test set of SWE-Bench and RepoQA. To ensure the repositories are sufficiently active and could provide a substantial number of resolved PRs, we filter out repositories with fewer than 1,000 stars or fewer than 1,000 Pull Requets (PRs). In the end, we obtain 229 open-source repositories.

\textbf{Issues and PRs Scraping}. For the selected repositories, we scrape their resolved issues and the corresponding PRs to construct the ReSAT training data. We utilze \textit{GhApi}~\footnote{https://ghapi.fast.ai} to scrape all the PRs in the repositories. Given a PR, we use regular expressions to extract the issue numbers mentioned in the PR title and commit messages, which are considered as the issues resolved by the PR. We filter out PRs that are not merged to the main branch and those do not mention any issues. As a result, we obtain 44,088 PRs in total to construct the ReSAT dataset.

\subsection{Localization Data Construction}

When resolving an issue, the most challenging part is accurately locating the code snippet that requires modification within a large repository. As shown in Figure ~\ref{fig:approach}(2), inspired by the design of the localization module in previous work~\cite{xia2024agentless,jimenez2024swebench,yang2024swe}, we construct a multi-level localization dataset:
(1) \textbf{File localization}: Given an issue and the repository file structure, identify the files that require to be modified to resolve the issue.
(2) \textbf{Function localization}: Given an issue and the modified file skeleton (including the class and function declarations), locate the classes and functions relevant to resolving the issue.
(3) \textbf{Line localization}: Given an issue and the full class or function content, locate the exact lines of code that need to be modified.

\subsubsection{File Localization}

File localization training enhances the SLMs' understanding of the high-level architecture of the repository, enabling it to perform an initial rough localization based on the issue description. We first clone the repository locally, then employ $os.walk$~\footnote{https://docs.python.org/3/library/os.html\#os.walk} to extract the repository file structure.
We utilize the issue and repository structure as inputs, with the full names of the modified files in the PR as outputs. To improve the quality of the training data, we excluded non-Python files and test scripts from both the repository structure in the input and the filenames in the output.
Template \ref{exmp:file_localization} shows the data template~\footnote{Due to space constraints, the prompt presented here has been simplified.} used for file localization training.

\subsubsection{Function Localization}
Function localization training can improve the SLMs' performance on fine-grained localization based on the functional characteristics of the code. After locating the files to be modified, further function-level localization is required due to the potentially large file contents. We employ $libcst$~\footnote{https://github.com/Instagram/LibCST}, a Concrete Syntax Tree (CST) parser, to extract class and function declarations from the file. In function localization, the issue and file skeleton composed of function names are employed as inputs, while the names of modified functions from the PRs are employed as outputs.
Template \ref{exmp:function_localization} shows the data template used for function localization training.

\subsubsection{Line Localization}
Line localization training enhances the SLMs' ability to accurately locate the specific lines of code that require to be modified to resolve the issue. In earlier stages, the identified functions may still be relatively long. Moreover, the file and function localization phases are based only on the project and file structure, lacking detailed code content information. We employ line localization training to correct the errors in function localization caused by missing information and leverage code details for precise locating. Specifically, we extract the modified lines in the PR and their corresponding functions. Line localization takes the issue description and function content as inputs and outputs the modified lines from the PR. 
Template \ref{exmp:line_localization} shows the data template used for line localization training.

\subsection{Code Edit Data Construction}
Code Edit  training can enhance the SLMs' ability to modify code snippets based on the issue. The input for Code Edit  consists of the issue and the localized code snippet, while the output is the code edits of the corresponding PR. Following previous work~\cite{xia2024agentless, yang2024swe}, we employ the \textit{Search/Replace} Edit format. The \textit{Search/Replace} format consists of two main parts: 1) Search: the original code snippet that need to modify, and 2) Replace: the new code snippet after editing. Compared to directly generating edits in diff format, the \textit{Search/Replace} format is easier to generate, and can be converted into the diff format through post-processing (e.g., employing \textit{difflib}~\footnote{https://docs.python.org/3/library/difflib.html}).
Template \ref{exmp:edit_generation} shows the data template used for code edit training.

\begin{table*}
    \centering
    \caption{Performance comparison on SWE-Bench-verified. ReSAT-$\ast$ is the model trained on ReSAT data. 
    }
    \label{tab:RQ1}
    \resizebox{.85\linewidth}{!}{
        \begin{tabular}{c|c|c|ccc}
        \hline
        \multirow{2}{*}{Framework} & \multirow{2}{*}{Model} & \multirow{2}{*}{\% Resolved} & \multicolumn{3}{c}{Localization}  \\
        \cline{4-6}
        &&& \% FileHit & \% FuncHit & \% LineHit \\
        \hline
        \multirow{4}{*}{RAG-SWE} 
        &Deepseek-Coder & 0.8 & 24.8 & - & - \\
        &   ReSAT-Deepseek-Coder (ours) &  \textbf{2.6} &  24.8 & \textbf{-} & \textbf{-} \\
        &CodeQwen & 1.4 & 24.8 & - & - \\
        &   ReSAT-CodeQwen (ours) &  \textbf{2.0} &  24.8 & \textbf{-} & \textbf{-} \\
        
        \hline
        \multirow{4}{*}{Agentless} 
        &Deepseek-Coder & 1.8 & 42.2 & 19.0 & 6.0 \\
        &   ReSAT-Deepseek-Coder (ours) &  \textbf{6.6} & \textbf{46.8} & \textbf{42.2} & \textbf{16.8} \\
        &CodeQwen & 0.8 & 51.4 & 25.8 & 9.0 \\
        &   ReSAT-CodeQwen (ours) &  \textbf{7.2} & \textbf{53.4} & \textbf{48.0} & \textbf{16.8} \\
        
        \hline

        \end{tabular}
    }
\end{table*}

\subsection{Training Details}

To further improve the quality of the training data, we filter out PRs that do not modify Python files and samples with a context length greater than 32k tokens. In the end, we construct 80,260 training samples from the 229 open-source repositories. We fine-tune CodeQwen1.5-7B-Chat and Deepseek-Coder-6.7B-Instruct through FastChat~\cite{zheng2023judging} framework. We set the max length of $tokenizer$ for both models as 32k, and apply linear rope scaling to Deepseek-Coder-6.7B-Instruct to scale up its max length.
The training process is conducted on 8x 80G A100 GPUs with full sharding strategy and CPU offload strategy implemented by Pytorch FSDP~\footnote{https://pytorch.org/docs/stable/fsdp.html}. We also utilize flash-attention-2~\cite{dao2023flashattention2} to reduce memory overhead and speed up the training process. We set the global batch size to 128 and train for 2 epochs. We apply cosine learning rate decay with a maximum learning rate of 5e-6 and 3\% warm-up steps. The entire training process takes about 11 hours for each model.

\section{Evaluation}\label{sec:setup}





\subsection{Datasets}

To evaluate the effectiveness of ReSAT on repository-level code understanding and issue resolving, our experiments are mainly conducted on two datasets: RepoQA and SWE-Bench-verified.

\textbf{RepoQA}~\cite{liu2024repoqa} is a benchmark designed to evaluate the ability of LMs to understand repository-level long-context code. The benchmark consists of a long-context code repository and a function description. LMs are required to find the function that matches the description within the repository and output the complete function. RepoQA is constructed by 500 long-context test samples across 5 programming languages: \textit{python}, \textit{c++}, \textit{rust}, \textit{java} and \textit{typescript}. 

\textbf{SWE-Bench-verified}~\cite{swebench-verified}, a manually verified issue resolving test dataset, is jointly constructed by OpenAI and the SWE-Bench team. Each issue is associated with test cases that can be executed in a Docker environment. SWE-Bench-verified requires LMs to edit the repository and pass all test cases, providing a more reliable evaluation of the model's issue resolving capabilities. SWE-Bench-verified consists of 500 high-quality test samples.

SWE-Bench-verified is utilized to evaluate ReSAT-trained SLMs' issue resolving capabilities, while RepoQA reflects the repository-level code understanding ability.

\subsection{Models}

In our experiments, we apply ReSAT training to two open-source code SLMs: Deepseek-Coder-6.7B-Instruct and CodeQwen1.5-7B-Chat. 

\textbf{Deepseek-Coder}~\cite{guo2024deepseek}. We employ Deepseek-Coder-6.7B-Instruct, a code SLM released by the deepseek team. 
It has demonstrated impressive performance on basic code generation tasks, with 78.6\% accuracy on HumanEval and 73.2\% accuracy on MBPP. Deepseek-Coder model is pre-trained on repository-level code corpus by employing a window length of 16k and an extra \textit{fill-in-the-blank}~\cite{guo2024deepseek} task, making it support repository-level code completion and infilling. But it still performs poorly on repository-level code understanding and issue resolving tasks, successfully resolving only 0.22\% issues in the original SWE-Bench.

\textbf{CodeQwen}~\cite{codeqwen1.5}. We employ CodeQwen1.5-7B-Chat, a specialized code SLM built upon the Qwen1.5-7B language model. 
It has outperformed larger models in basic code generation tasks, achieving 83.5\% accuracy on HumanEval and 77.7\% accuracy on MBPP. It possesses the capability to understand and generate long-context code with up to 64k tokens, but its performance still remains suboptimal on repository-level tasks, successfully resolving only 0.89\% issues in the original SWE-Bench.


\subsection{Inference Framework}
To directly reflect the ability of ReSAT-trained SLMs to solve issues in real-world applications, we apply ReSAT-trained SLMs on two existing issue resolving frameworks.

\textbf{Agentless}~\cite{xia2024agentless}, a straightforward LM-based framework for repository-level issue resolving. It contains two parts: (1) Localization: LMs are utilized to identify the code snippets that are responsible for the issue by hierarchically narrowing down candidate files, classes, functions, and even lines of code. (2) Code edit generation: LMs are required to generate potential edits to resolve the issue. 

\textbf{RAG-SWE}~\cite{jimenez2024swebench}, first retrieves the files that are most similar to the issue description, then uses the file content and the issue as prompt for LMs to generate code edits. Following previous work~\cite{jimenez2024swebench}, we utilize BM-25 as our retriever, and retrieve the top-3 most similar files from the repositories. Then we apply the same generation phase as Agentless.


\subsection{Metrics}
When we evaluate issue resolving frameworks with ReSAT-trained SLMs on SWE-Bench-verified dataset, to fairly and thoroughly evaluate impact of ReSAT on issue resolving, we apply four metrics: \textbf{\% Resolved}, \textbf{\% FileHit}, \textbf{\% FuncHit}, and \textbf{\% LineHit}. 
\textbf{\% Resolved} is the proportion of test samples that the issue resolving frameworks can successfully generate code edits based on issues and pass all test cases.
\textbf{\% FileHit} refers to the proportion of files modified in the PRs that are successfully predicted by the SLMs during file localization.
\textbf{\% FuncHit} refers to the proportion of functions modified in the PRs that are successfully predicted by the SLMs during Function localization.
\textbf{\% LineHit} refers to the proportion of lines modified in the PRs that are successfully predicted by the SLMs during line localization.

When we evaluate ReSAT-trained SLMs on RepoQA dataset, we apply  the same \textbf{Accuracy} metric as the original evaluation setting in RepoQA~\cite{liu2024repoqa}, which refers to the proportion of LMs predicting the correct outputs for the samples.


\begin{table*}
    \centering
    \caption{Performance comparison on RepoQA. The middle five columns are the accuracy on single programming language, the rightest column is the average accuracy across five lanuages.}
    \label{tab:RQ2}
    \resizebox{.75\linewidth}{!}{
        \begin{tabular}{c|ccccc|c}
        \hline
         \multirow{2}{*}{Model} & \multicolumn{5}{c|}{Single-Lanuage Accuracy} & \multirow{2}{*}{Avg. Accuracy}  \\
         \cline{2-6}
        & .py & .cpp & .rs & .java & .ts & \\ 
        \hline
        CodeQwen & 69 & 47 & 47 & 74 & \textbf{67} & 62.8 \\
         ReSAT-CodeQwen (ours) &  \textbf{73} & \textbf{52} & \textbf{61} & \textbf{76} &  65 & \textbf{65.4}\\
        \hline
        Deepseek-Coder & 11 & \textbf{21} & 2 & 3 & 16 & 10.6 \\
         ReSAT-Deepseek-Coder (ours) &   \textbf{13} &  14 & \textbf{14} & \textbf{13} & \textbf{21} & \textbf{15.0}\\
        \hline
        
        \end{tabular}
    }
\end{table*}



    

\begin{figure*}[t]
    \centering
    \includegraphics[width=.99\textwidth]{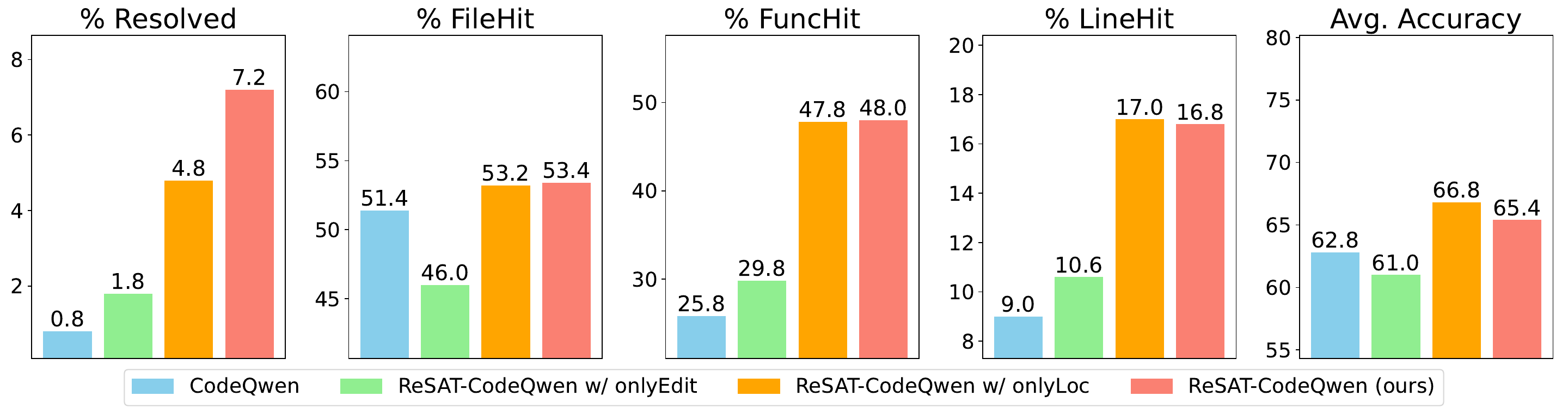}
    \caption{
    Ablation study on training datasets. The first four figures show the metrics on SWE-bench-verified. The last one shows the average accuracy on RepoQA.
    }
    \label{fig:ablation}
\end{figure*}

\subsection{Main Results}


\paragraph{ReSAT enhances issue resolving performance.}
Table \ref{tab:RQ1} shows the evaluation results on SWE-Bench-verified before and after ReSAT training. For example, when employing the Agentless framework, ReSAT training improves the \% Resolved for Deepseek-Coder and CodeQwen by 4.8\% and 6.4\%. ReSAT training also improves the File, Function, and Line-level localization performance. When employing the Agentless framework, ReSAT improves Deepseek-Coder's \%FileHit, \%FuncHit, and \%LineHit by 4.6\%, 23.2\%, and 10.8\%. 
The \%FileHit performance gap between RAG-SWE and Agentless with ReSAT-CodeQwen (24.8\% v.s 53.4\%), also demonstrates that employing ReSAT-trained SLMs for localization yields higher accuracy, which further leads to Agentless successfully resolving more issues than RAG-SWE after ReSAT training.


\paragraph{ReSAT enhances repository-level code understanding performance.}
As shown in Table \ref{tab:RQ2}, after ReSAT training, both models demonstrate improved performance on RepoQA. CodeQwen's average accuracy increases from 62.8 to 65.4, and Deepseek-Coder's average accuracy improves from 10.6 to 15.0. The results in Table \ref{tab:RQ2} also indicate that training exclusively on ReSAT data in Python can still improve performance in other languages. After ReSAT training, SLMs achieve higher accuracy in most languages.


\subsubsection{Ablation Study}

To explore the impact of ReSAT's localization and code edit data on repository-level code understanding and issue-resolving capabilities, we conduct ablation experiments on ReSAT's training data using CodeQwen. 

Figure \ref{fig:ablation} shows the evaluation results of CodeQwen trained on single dataset. The results on SWE-Bench-verified indicate that \textbf{both types of training data have a positive effect on the issue-resolving capability of SLMs}. Compared to original CodeQwen, training with only code edit data and localization data improves the \% Resolved by 1.0 and 4.0.
Combining both types of data results in stronger issue-resolving capabilities than using single data, using only code edit data and localization data reduced the \% Resolved by 5.4 and 2.4. 
Localization Hits on SWE-Bench-verified and Accuracy on RepoQA indicate that localization data effectively enhances SLMs' repository-level code understanding. 
\textbf{Localization data has greater impact on issue-resolving performance compared to code edit data}, further supporting the hypothesis that training focused on repository structure understanding can effectively improve issue-resolving capabilities.





\begin{figure*}
    \centering
    \subfloat[Successful example.]{
    \includegraphics[width=0.48\textwidth]{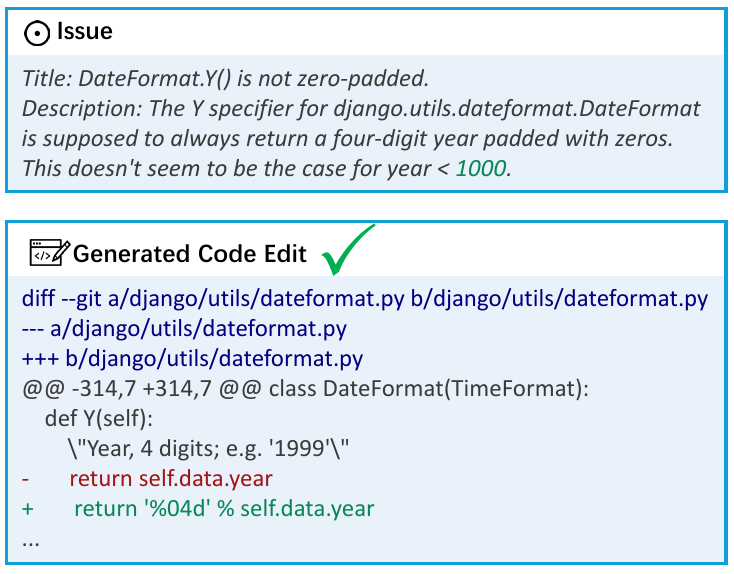}}\quad
    \subfloat[Failed example.]{
    \includegraphics[width=0.48\textwidth]{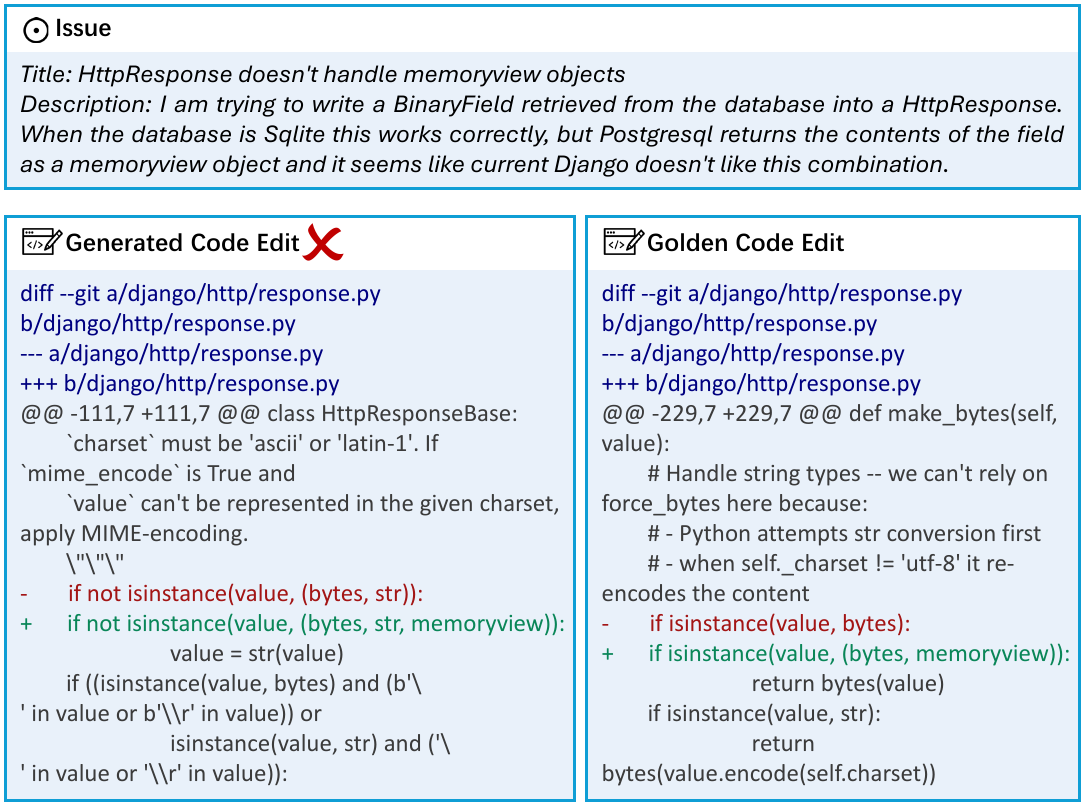}}\quad
    \caption{Case study.}
    \label{fig:case}
\end{figure*}

\subsection{Case Study}
We illustrate the issue-resolving performance of ReSAT-trained CodeQwen through two examples. In the first example of Figure \ref{fig:case}, the issue requires adjusting the format of \textit{"year"} to four digits. ReSAT-trained CodeQwen successfully identifies the necessary function $Y$ and modifies the format of \textit{"year"} to four digits. In the second example of Figure \ref{fig:case}, ReSAT-trained CodeQwen successfully understands that the issue requires checking whether the value is a $memoryview$, but it modifies the wrong location. The failure may be due to the issue being expressed indirectly, requiring a more complex logical reasoning process. In future work, constructing \textit{Chain-Of-Thought}~\cite{wei2022chain} style training data may further improve SLMs' reasoning capabilities of resolving issues.

\section{Related Work}\label{sec:related_work}

\subsection{Issue Resolving with LMs.}

The rapid development of language models (LMs) has made automated issue resolving possible. There are two types of automatic issue resolving systems: agent-based and pipeline-based. 

Agent-based systems~\cite{wu2023autogen, yang2024swe, chen2024coder, hong2023metagpt, ishibashi2024self, luo2024repoagent} equip LMs with tools for decision-making and iterative actions. \citet{yang2024swe} proposed the SWE-Agent and designed an Agent-Computer Interface (ACI) that is more suitable for LMs than IDEs, allowing LMs to automatically invoke these ACIs to edit files, navigate repositories, and execute tests. \citet{chen2024coder} proposed CodeR, which distributes tasks to different agents based on a task graph, addressing issues through multi-agent collaborations. 

Pipeline-based systems~\cite{xia2024agentless, jimenez2024swebench, liang2024repofuse, shrivastava2023repository} follow a streamlined two-phase pipeline of localization and generation. 
\citet{jimenez2024swebench} proposed RAG-SWE system for issue resolving, which first retrieves the files most similar to the issue description. Then, the issue and the retrieved files are used together as prompt for the LMs to generate code edits. 
\citet{xia2024agentless} proposed Agentless, a system that developed to automate software development tasks without the complexity of autonomous agents. The LMs are responsible for identifying the location of the issue within the repository and generating candidate edits, and it does not autonomously decide future actions or rely on complex tools. 

\subsection{Training-Data Synthesis for LMs.}

After the pre-training of LMs, a large amount of human-annotated question-answer data is required for instruction tuning, which incurs significant human labor costs. Some work has attempted to generate training data through data synthesis. \citet{wang2022self} proposed Self-Instruct, which maintains an instruction pool. Instruction examples are randomly selected from the pool, and LMs are prompted to generate new instructions based on the examples, thereby continuously expanding the diversity of instructions in the training data. \citet{xu2024wizardlm} proposed Evolve-Instruct, which enhances the diversity and complexity of instructions by evolving existing instructions in terms of difficulty, domain, and other factors.

Some work has focused on synthesizing domain-specific training data to enhance LMs' specialized capabilities~\cite{luo2023wizardcoder, wei2024magicoder, an2023learning, muennighoff2023octopack}. ~\citet{luo2023wizardcoder} proposed Code Evolve-Instruct, which increases the complexity and difficulty of synthesized coding tasks through instruction evolution. ~\citet{wei2024magicoder} introduced OSS-Instruct, incorporating code snippets from open-source communities into the instruction evolution process to enhance the diversity of synthesized data. All of the above work relies on powerful LLMs to synthesize training data. In this paper, we propose ReSAT, instead of relying on LLMs to synthesize data, directly utilizes open-source repositories to synthesize training data.


\section{Conclusion}\label{sec:conclusion}
In conclusion, we propose ReSAT, a repository structure-aware training approach.
ReSAT constructs two types of training data: localization data and code edit data. 
Evaluation results on SWE-Bench-verified and RepoQA demonstrate that ReSAT effectively enhances SLMs' issue-resolving and repository-level long-context understanding capabilities. 
In future work, we will further expand the scale of ReSAT data and improve the efficiency of leveraging open-source repositories, continuing to enhance the performance of SLMs-based automatic programming assistants.

\section{Limitations} \label{sec:limitation}

Due to limitations in computational resources, we only conducted training on two models with 7 billion parameters. We believe that ReSAT can also enhance the performance of other LMs with different parameter sizes on repository-level tasks, we leave it for the furture work.

Since SWE-Bench only consists of Python issue-resolving tasks, our experiments only verified that ReSAT improves issue-resolving capabilities for Python repositories. Given that projects in other programming languages can also provide the structural information required by ReSAT, we believe that ReSAT can improve issue-resolving performance in other languages. Moreover, our results on RepoQA also show that ReSAT training with Python improves performance in other languages.

Despite the effectiveness of ReSAT in enhancing the issue-resolving capabilities of SLMs, there remains a significant gap compared to LLMs like GPT-4o. Future work should further enhance the repository structure understanding and code editing capabilities of SLMs to bridge this gap.

ReSAT training requires a certain amount of computational resources. Considering the improvement ReSAT brings to the issue-resolving capabilities of SLMs, we believe this consumption is worthwhile. Future work should focus on more environmentally-friendly research, exploring how to achieve improvements in issue-resolving capabilities with less computational resource consumption.

\bibliography{reference}

\appendix
\clearpage
This is the Appendix of the paper: \textit{Repository Structure-Aware Training Makes SLMs Better Issue Resolver}.

\begin{figure*}[h]
    \centering
    \includegraphics[width=0.8\textwidth]{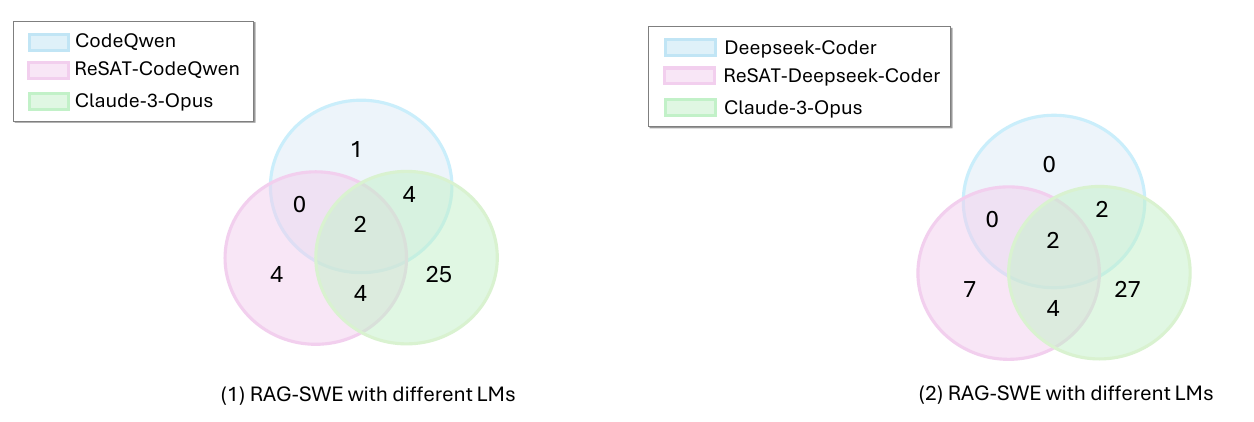}
    \caption{Comparison results between RAG-SWE with ReSAT-trained SLMs and LLMs.}
    \label{fig:SLMsVSLLMs}
\end{figure*}

\section{Compare ReSAT-trained SLMs with Advanced LLMs}
We evaluate the performance of SLMs and advanced LLMs on SWE-Bench-verified using the same programming assistants. The results in Table ~\ref{tab:SLMsVsLLMs} show that \textbf{ReSAT training has effectively narrowed the gap between SLMs and advanced LLMs in issue-resolving}. Under the RAG-SWE framework, the \% Resolved for ReSAT-trained Deepseek-Coder increases from 0.8 to 2.6, leaving only 0.2 behind GPT-4.

To further understand the impact of ReSAT training on the issue-resolving capabilities of SLMs, we compare the issues resolved by ReSAT-trained SLMs and Claude-3-Opus when applied to the RAG-SWE framework. The results in Figure ~\ref{fig:SLMsVSLLMs} show that ReSAT-CodeQwen successfully solves 4 issues that Claude failed, and ReSAT-Deepseek-Coder successfully solves 7 issues that Claude failed. The comparison indicates that \textbf{ReSAT training enables SLMs to solve issues that even advanced LLMs failed}, making the combination of SLMs and LLMs a viable solution for more effectively issue resolving.

\begin{table}[h]
    \centering
    \caption{Compared ReSAT-trained SLMs with advanced LLMs.}
    \label{tab:SLMsVsLLMs}
    \resizebox{.9\linewidth}{!}{
        \begin{tabular}{c|c|c}
        \hline
        Framework & Model & \% Resolved \\
        \hline
        \multirow{6}{*}{RAG-SWE} & CodeQwen & 1.4\\
        &  ReSAT-CodeQwen (ours) &  2.0 \\
        \cline{2-3}
        & Deepseek-Coder & 0.8 \\
        &  ReSAT-DeepSeek-Coder (ours) &  2.6 \\
        \cline{2-3}
        & GPT-4 & 2.8 \\
        & Claude-3-Opus & \textbf{7.0} \\
        \hline
        \multirow{5}{*}{Agentless} & CodeQwen & 0.8 \\
        &  ReSAT-CodeQwen (ours) &   7.2 \\
        \cline{2-3}
        & DeepSeek-Coder & 1.8 \\
        &  ReSAT-DeepSeek-Coder (ours) &  6.6 \\
        \cline{2-3}
        & GPT-4o & \textbf{33.2} \\    
        \hline
        \end{tabular}
    }
\end{table}

\section{Compare ReSAT with Other Training Approaches}
Some previous work also  attempts to construct training data for specific issue resolving frameworks to improve the issue-resolving capabilities of SLMs. To further validate the effectiveness of ReSAT, we conduct comparison between Agentless with ReSAT-trained CodeQwen and two other issue resolving frameworks with trained SLMs:
\begin{itemize}
    \item \textbf{Opendevin with CodeQwen-Opendevin}~\cite{wang2024opendevin}: Opendevin is an open-source agent-based issue resolving framework, while CodeQwen-Opendevin is the CodeQwen model trained on data specifically constructed for Opendevin.
    \item \textbf{RAG-SWE with SWE-Llama}~\cite{jimenez2024swebench}: RAG-SWE is the issue resolving framework used in RQ1, while SWE-Llama is the CodeLlama-Python-7B~\cite{roziere2023code} model trained on data specifically constructed for RAG-SWE.
\end{itemize}

\begin{table}
    \centering
    \caption{Comparison between Agentless with ReSAT-trained CodeQwen and two other issue resolving frameworks with trained SLMs.}
    \label{tab:finding4}
    \resizebox{.95\linewidth}{!}{
        \begin{tabular}{ccc}
        \hline
        Framework & Model & \% Resolved \\
        \hline
        Opendevin   & CodeQwen-Opendevin & 1.6 \\
        RAG-SWE     & SWE-Llama & 1.4 \\
         Agentless   &  ReSAT-CodeQwen (ours) &  \textbf{7.2} \\
        \hline
        
        \end{tabular}
    }
\end{table}

Table ~\ref{tab:finding4} shows the comparison results between Agentless with ReSAT-trained CodeQwen and two previous issue resolving frameworks with trained SLMs. The results show that Agentless with ReSAT-trained CodeQwen is able to resolve more issues. Agentless with ReSAT-trained CodeQwen resolves 7.2\% issues, while the other issue resolving frameworks with trained SLMs only resolves 1.4\% and 1.6\% issues. The comparison results demonstrate that the combination of ReSAT training and Agentless is the most effective approach for applying SLMs to issue resolving.

\section{Data Generation Prompt}
In this section we present the prompt template for file localization, function localization, line localization and code edit. 

Template \ref{exmp:file_localization} shows the \textbf{file localization} prompt, the input is the issue and the repository's file directory, and the output is the files modified by the corresponding PRs, which helps enhance the SLMs' repository structure understanding and coarse-grained localization ability.

\begin{exmp}{\textit{File Localization} Template}{file_localization}

\textbf{Input:}\\
Please look through the following GitHub problem description and Repository structure and provide a list of files that one would need to edit to fix the problem.

\textit{Problem Description:}
\{ problem statement \}

\textit{Repository Structure:}
\{ structure \}

\textbf{Output:}\\
\textit{Localized Files:}
\{ localized files \}
\end{exmp}

Template \ref{exmp:function_localization} shows the \textbf{function localization} prompt, the input is the issue and the content of the files modified by the corresponding PR, and the output is the name of the functions modified by the PR, which improves the SLMs' ability to locate functions within the file. 

\begin{exmp}{\textit{Function Localization} Template}{function_localization}
\textbf{Input:}\\
Please look through the following GitHub Problem Description and the Skeleton of Relevant Files.
Identify all locations that need inspection or editing to fix the problem, including directly related areas as well as any potentially related global variables, functions, and classes.

\textit{Problem Description:}
\{ problem statement \}

\textit{Skeleton of Relevant Files:}
\{ file skeleton \}


\textbf{Output:}\\
\textit{Localized results:}
\{ localized functions \}
\end{exmp}

Template \ref{exmp:line_localization} shows the \textbf{line localization} prompt, the input is the issue and the code context modified by the corresponding PR, and the output is the specific lines modified by the PR, aims to enhance the SLMs' precise localization ability. Notably, to increase the complexity of the line localization task, we introduce irrelevant functions as distractions. We also include the actual modified function name from the PR as part of the output, to enhance SLMs' ability to accurately locate functions based on their content and relevance to the issue. This allows SLMs to filter out irrelevant functions during the line localization stage and correct errors from the function localization stage in practical applications.

\begin{exmp}{\textit{Line Localization} Template}{line_localization}
\textbf{Input:}\\
Please review the following GitHub problem description and relevant files, and provide a set of locations that need to be edited to fix the issue.
The locations can be specified as class names, function or method names, or exact line numbers that require modification.

\textit{Problem Description:}
\{ problem statement \}

\textit{File Contents:}
\{ file content \}

Please provide the class name, function or method name, or the exact line numbers that need to be edited.

\textbf{Output:}\\
\textit{Localized results:}
\{ localized functions and lines \}
\end{exmp}

Template \ref{exmp:edit_generation} shows the \textbf{code edit} prompt, the input is the issue and a code snippet, and the output is a \textit{Search/Replace} style code edit.

\begin{exmp}{\textit{Code Edit} Template}{edit_generation}
\textbf{Input:}\\
You will be provided with an issue statement explaining a problem to resolve and a partial code base. Please first localize the bug based on the issue statement, and then generate *SEARCH/REPLACE* edits to fix the issue.

\textit{Problem Description:}
\{ problem statement \}

\textit{File Contents:}
\{ file content \}

Please first localize the bug based on the issue statement, and then generate *SEARCH/REPLACE* edits to fix the issue.

\textbf{Output:}\\
\textit{Code Edits:}
\{ code edits \}
\end{exmp}

\end{document}